\documentclass{optica-article}

\journal{opticajournal}

\articletype{Research Article}
\usepackage{lineno}
\usepackage{xcolor}
\begin{document}

\title{Optical transmitter for time-bin encoding Quantum Key Distribution}

\author{Julián Morales,\authormark{1,2,*} M. Guadalupe Aparicio,\authormark{2} Carlos F. Longo,\authormark{4} Cristian L. Arrieta,\authormark{4} and Miguel A. Larotonda\authormark{1,2,3}}

\address{\authormark{1}UNIDEF, Ministerio de Defensa-CONICET, J.B. de Lasalle 4397, 1603 Villa Martelli, Buenos Aires, Argentina\\
\authormark{2}Departamento de Física, Facultad de Ciencias Exactas y Naturales, Universidad de Buenos Aires. Ciudad Universitaria Pabellón I, 1428 Buenos Aires, Argentina\\
\authormark{3}Departamento de Investigaciones en Láseres y Aplicaciones, CITEDEF, Ministerio de Defensa, J.B. de Lasalle 4397, 1603 Villa Martelli, Buenos Aires, Argentina\\
\authormark{4}Departamento de Electrónica Aplicada, CITEDEF, Ministerio de Defensa, J.B. de Lasalle 4397, 1603 Villa Martelli, Buenos Aires, Argentina}

\email{\authormark{*}jmorales@citedef.gob.ar} %% email address is required; see note below about the corresponding author designation
%%julian95morales@gmail.com
% \homepage{http:...} %% author's URL, if desired

%%%%%%%%%%%%%%%%%%% abstract %%%%%%%%%%%%%%%%
%% [use \begin{abstract*}...\end{abstract*} if exempt from copyright]

\begin{abstract}
We introduce an electro-optical arrangement that is able to produce time-bin encoded symbols with the decoy state method over a standard optical fiber in the C-band telecom window. The device consists of a specifically designed pulse pattern generator for pulse production, a field-programmable gate array that controls timing and synchronization. The electrical pulse output drive a sequence of intensity modulators acting on a continuous laser that deliver bursts of weak optical pulse pairs of discrete intensity values. Such transmitter allows for the generation of all the quantum states needed to implement a discrete variable Quantum Key Distribution protocol over a single-mode fiber channel. Symbols are structured in bursts; the minimum relative delay between pulses is 1.25~ns, and the maximum symbol rate within a burst is 200 MHz. We test the transmitter on simulated optical channels of 7~dB and 14~dB loss, obtaining maximum extractable secure key rates of 3.0~kb/s and 0.57~kb/s respectively. Time bin state parameters such as symbol rate, pulse separation and intensity ratio between signal and decoy states can be easily accessed and changed, allowing the transmitter to adapt to different experimental conditions and contributing to standardization of QKD implementations.
\end{abstract}

%%%%%%%%%%%%%%%%%%%%%%%%%%  body  %%%%%%%%%%%%%%%%%%%%%%%%%%
\section{Introduction}
\label{sec:intro}

Quantum Key Distribution (QKD) is a cryptographic task that relies on fundamental principles of quantum mechanics that allows for two parties to share a random secret key. The presence of an eavesdropper trying to gain information on the key disturbs the system and introduces measurable changes that reveals their presence  \cite{gisin2002quantum}.

The first protocol dates back to 1984, when Charles Bennett and Gilles Brassard introduced the BB84 protocol \cite{bennet1984quantum} and shortly after the first experimental realization was demonstrated \cite{bennett1992experimental}. Since then, many protocols, techniques and emerging technologies have contributed to bolster this rapidly expanding field. Preferred channels for these quantum communications protocols are optical fibers in the near infrared spectral region \cite{wang2014field,zhang2019quantum,mao2018integrating,chen2021twin} and, most recently, open space links between ground stations and low earth orbit satellites \cite{liao2017satellite,ren2022portable,lu2022micius}. Ground based QKD protocols have evolved to simpler and more efficient schemes combined with robust security that feature symbol rates in the order range of the GHz \cite{takesue2007quantum,wang20122,shibata2014quantum,dynes2016ultra,grunenfelder2018simple,boaron2018simple}, and record length transmission links exceeding 400 km of single mode fibers \cite{boaron2018secure} for prepare-and-measure protocols, 800 km for the novel twin-field protocols \cite{wang2022twin} and 1000 km for entanglement-based protocols \cite{yin2020entanglement} . Meanwhile, quantum communications and QKD have been demonstrated on satellite-to-earth open space links, using polarization encoding \cite{liao2017satellite}.
Despite all these achievements and developments, QKD is still not a mature technology: Although solutions that can perform the tasks on both the emitter and the receiver sides in an efficient way do exist, these building blocks for a quantum communication system are still not standardized. Only a few companies offer commercial (yet closed) systems for quantum secure communications.  
In order to obtain a fast and efficient QKD implementation, a few key aspects have to be considered: a fast and reliable state preparation, a simplified experimental realization with minimum technical requirements and its security guaranteed by quantum physics,  combined with an efficient detection scheme. 

For fiber based links, the preferred qubit encoding is time-bin (arrival time of the detected photons, relative to a fixed clock reference). In general, the most easily implementable protocols (i.e. the ones that require minimum active parts and processes) are also the ones that achieve highest speeds. Furthermore, they have the advantage of being less prone to side channel attacks, since almost every imperfection of an optical component might be exploited as a security weakness \cite{boaron2018simple,boaron2018secure,namekata2011high}. Modern protocols such as Measurement-device-independent QKD \cite{lo2012measurement,tang2014experimental,yin2016measurement} require spectral, temporal and polarization indistinguishability of signal pulses generated by the two independent and distant laser sources. On the other hand, twin-field QKD \cite{lucamarini2018overcoming,wang2018twin,cui2019twin,chen2020sending} adds the need for link phase stabilization. The respective increased security and distance between nodes that these protocols offer come at the cost of an increased experimental complexity, although much effort is being put into these technologies \cite{woodward2021gigahertz,clivati2022coherent}. Regarding the detection, fast and efficient sensing requirements have lead to alternative technologies such as superconducting nanowire single photon detectors \cite{gol2001picosecond,marsili2013detecting}, or self-differencing techniques that increase the gating frequency of traditional avalanche photodiodes \cite{yuan2007high,dixon2008gigahertz}. Also, minimizing the temporal multiplexing of auxiliary signals or processes such as active interferometer stabilization can increase the maximum detection rate. 

On the other side of the link, the emitter should be able to prepare all the quantum states in a fast and unambiguous manner. This implies shaping pulses as symbols belonging to different state bases and the ability to prepare signal and decoy states. For practical encoding of time-bin qubits, defined by two temporal modes with a relative delay $\Delta T$ called early ($e$) and late ($l$), states of the $Z$ basis are approximated using weak coherent pulses of mean photon number $\mu=\left|\alpha\right|^2$:
\begin{equation}
 \begin{aligned}
 \label{eq:zstates}
  \left|\psi_0 \right \rangle&=\left|\alpha \right  \rangle_e \left|0 \right \rangle_l \\ 
  \left|\psi_1 \right \rangle&=\left|0 \right \rangle_e \left|\alpha \right \rangle_l 
 \end{aligned}
\end{equation}

States of the $X$ basis are superpositions of $\left|\psi_0 \right \rangle$ and $\left|\psi_1 \right \rangle$: $ \left|\psi_{\pm} \right \rangle=1/\sqrt{2}\left(\left|\psi_0 \right  \rangle \pm \left|\psi_1 \right  \rangle \right) $, and correspond to a superposition of two wavepackets temporally separated by $\Delta T$. Measurements in the $X$ basis require the use of an interferometer with an arm unbalance of $\Delta T$: states are split into the two arms and upon recombination, phase information can be retrieved from detections of the central temporal bin \cite{marand1995quantum,tittel1998experimental}. Thus, in order to avoid instabilities from large arm unbalance, and to be able to increase the symbol rate, the temporal bins should be set as close as possible, by minimizing $\Delta T$. This condition imposes a large RF bandwidth for the  optoelectronic components to shape the coding symbols, regardless of the specific method used to generate the pulses.

In this work we present a transmitter for a variable rate  QKD system over an optical fiber channel; an optical source that is capable of generating quantum signals for a discrete variable decoy state Quantum Key Distribution protocol with time-bin encoding. 
Due to their flexibility, multiple task managing and processing power, state-of-the-art designs use either high performance field programmable gate arrays (FPGAs) and peripherals, or RF arbitrary waveform generators to execute all the tasks required by the communication protocol \cite{korzh2013high,paraiso2019modulator,grande2021adaptable}. Following the former approach, we generate all the timing, triggering and synchronizing signals with a development board based on the Xilinx Zynq-7000 FPGA, while the fast pattern that is needed for optical pulse generation is produced with an ad-hoc high speed circuit.

The following section is devoted to the description of the electronics, opto-electronics and driving signals needed to implement a QKD protocol based on time-bin encoding on weak coherent pulses using a decoy state method \cite{hwang2003quantum,lo2005decoy}. We show the performance of the transmitter in section \ref{sec:perform}.

\section{Experimental arrangement}
\label{sec:expmt}

The devised transmitter is able to produce the necessary states to implement a three-state protocol \cite{tamaki2014loss,mizutani2015finite,boaron2018simple} with the one-decoy state method, that implies the generation of symbols with two different mean photon numbers, $\mu_1$ (signal) and $\mu_2$ (decoy). These states, encoded in the time-bin degree of freedom can be expressed as follows:
\begin{equation}
 \renewcommand*{\arraystretch}{1.5}
\begin{matrix}
\left|\psi_0 \right \rangle_{\mu_1}=\left|\sqrt{\mu_1} \right  \rangle \left|0 \right \rangle;\;   & \left|\psi_0 \right \rangle_{\mu_2}=\left|\sqrt{\mu_2} \right  \rangle \left|0 \right \rangle  \\

 \left|\psi_1 \right \rangle_{\mu_1}=\left|0 \right  \rangle \left|\sqrt{\mu_1} \right \rangle;\;   & \left|\psi_0 \right \rangle_{\mu_2}=\left|0 \right  \rangle \left|\sqrt{\mu_2} \right \rangle  \\

 \left|\psi_+ \right \rangle_{\mu_1}=1/\sqrt{2}\left( \left|\sqrt{\mu_1}  \right  \rangle \left|0\right \rangle +\left|0 \right  \rangle \left|\sqrt{\mu_1} \right \rangle \right );\;   & \left|\psi_+ \right \rangle_{\mu_2}=1/\sqrt{2}\left( \left|\sqrt{\mu_2}  \right  \rangle \left|0\right \rangle +\left|0 \right  \rangle \left|\sqrt{\mu_2} \right \rangle \right )  \\ 
 
\end{matrix}
\label{eq:estados}
\end{equation}

States from the computational $Z$ basis, Eq. (\ref{eq:zstates}) are used to generate the key, while states $\left|\psi_+ \right \rangle$ are used to estimate the information obtained by an eavesdropper. An additional requirement for any implementation of QKD with weak coherent state signals is that the phase of each symbol must be random in order to enhance the security of the protocol \cite{lo2005phase}. This single decoy state method has been shown to outperform the 2-decoy protocol for almost all experimental settings under the assumption of  finite key length \cite{rusca2018finite}.

\begin{figure}[h!]
\centering\includegraphics[width=8.5cm]{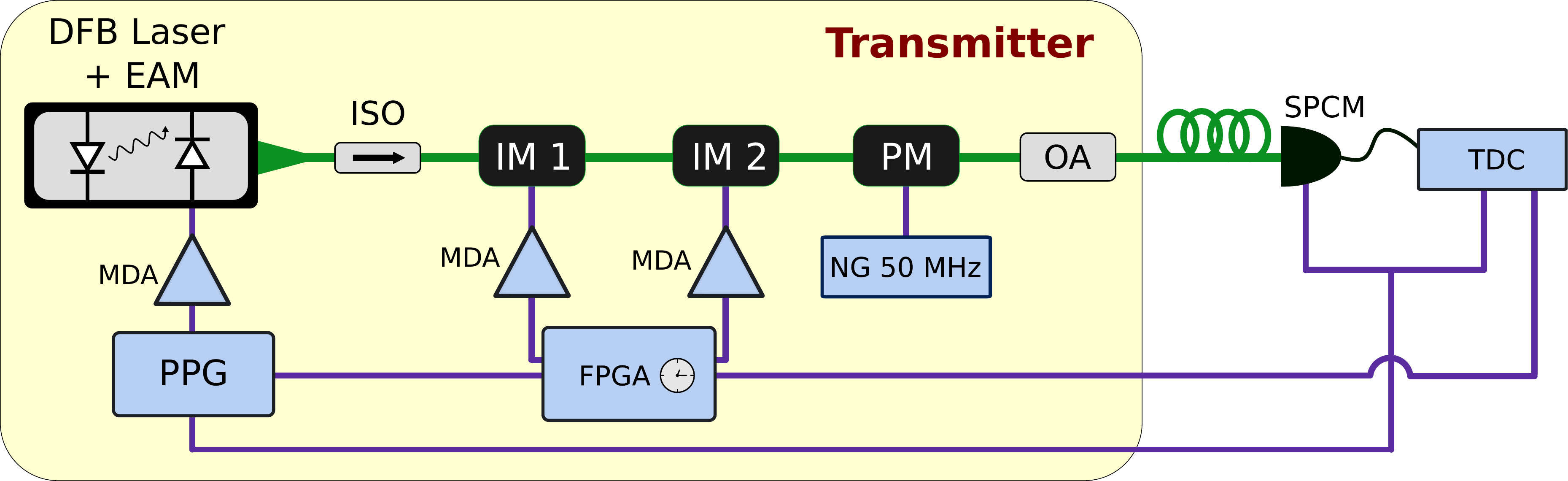}
\caption{Scheme of the proposed QKD transmitter. The EAM and the intensity modulators IM1 and IM2 that set the mean photon values for the $\left|\psi_+ \right \rangle$ states and the signal and decoy states respectively are driven by Modulator Driver Amplifiers (MDAs), a phase modulator (PM) driven by a noise generator (NG) randomizes the phase between symbols. ISO: optical isolator; OA: optical attenuator.  PPG refers to the Pulse Pattern Generator. Detection and state identification is done by means of a Single Photon Counting Module (SPCM) and a time-to-digital converter (TDC).}
\label{fig:Setup_b}
\end{figure}

The optical part of the transmitter consists of a 1548~nm continuous wave (CW) distributed feedback laser (Mitsubishi FU-641SEA) with an integrated electro-absorption modulator (EAM), and a chain of a phase modulator (Thorlabs LN65S-FC) and two intensity modulators (Lucent 2623NA) for phase randomization, normalization of the mean photon number of the  $\left|\psi_+ \right \rangle$ state and selection of signal or decoy states, respectively (Fig. \ref{fig:Setup_b}). A Digilent Zybo Zynq-7000 FPGA generates the main clock, the digital signals that drive the external intensity modulators and a synchronizing signal for detection. Since the bandwidth of the desired patterns is larger than the frequency range attainable by the FPGA digital outputs, optical pulses are carved out from the CW laser output using the 10 GHz built-in EAM, which is driven by a Pulse Pattern Generator (PPG) specifically designed for this task.

\begin{figure}[h!]
\centering\includegraphics[width=8.5cm]{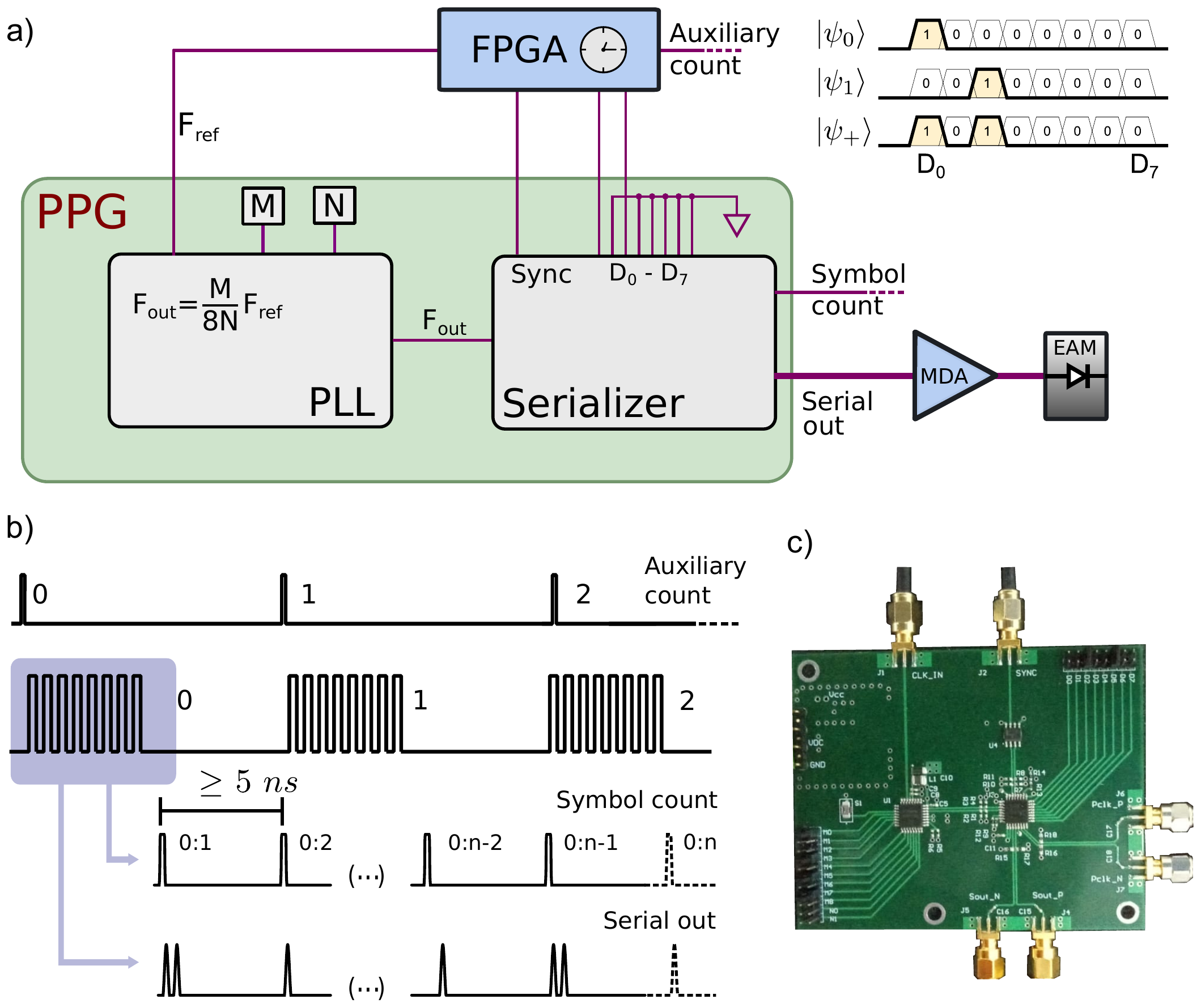}
\caption{a) Pulse Pattern Generator (PPG). A PLL clocking circuit synthesizes a tunable multiple ($F_{\text{out}}$) of the input frequency ($F_{\text{ref}}$), which serves as the serializer clock. The initial single or double pulse patterns are selected by setting the parallel bus ($D_0$-$D_7$) with the FPGA, and an additional input signal ($Sync$) enables the output of the PPG and allows for structuring of the data stream. Patterns with minimum delay between pulses are produced using the 8-bit sequences depicted on the top right; b) Timing diagram. The output is structured in symbol bursts. Each burst is composed by a user-defined amount of symbols, which can be produced at a maximum rate of 200 MHz. Auxiliary timing and counting signals are generated at the FPGA on each burst and sent to the receiver. A low jitter counting signal synchronized with each data symbol is generated on the GPP to identify the time-bin state of each symbol; c) GPP board showing ECL differential serial (bottom edge) and clock (right edge) outputs.}
\label{fig:ppg}
\end{figure}

The PPG was designed and built in-house upon a 4 layer printed circuit board. It is based on a PLL clocking circuit and a 3.2 Gb/s data rate serializer. This arrangement generates a selected optical pulse pattern by converting an 8-bit parallel bus into a serial stream by means of a MC10EP446 serializer integrated circuit (Onsemi), at twice the frequency of its clock. The PPG accepts a clock input signal between 10 and 100~MHz, which is delivered by the FPGA board. A NBC12430 programmable phase-locked loop clock generator (Onsemi) synthesizes a clock signal of frequency $F_{out}$ between 400 and 800~MHz from the input frequency to clock the serializer (Fig. \ref{fig:ppg}a). The generator also includes AC coupled differential outputs for the serial pattern and for the parallel clock. An additional input signal (\emph{Sync}) managed by the FPGA synchronously enables the serial output, allowing for the decoupling between the symbol repetition rate and the pulses width and delay, and also for burst data structuring. Single or double pulse patterns needed to form states from $Z$ and $X$ bases are generated by serializing 8-bit parallel strings; for the minimum pulse separation, these strings are 10000000, 00100000 and 10100000 or any shift of these patterns. A faster digital data transfer from the FPGA to the serializer is currently under development, that will allow to send two symbols per serial string. Larger delays between \emph{early} and \emph{late} pulses can be obtained by including more \emph{off} bits between them; larger interferometer unbalances on the detection side can be accommodated in this way. Additionally, two timing signals are sent to the receiver through the classical channel to adequately count and identify states within symbols (Fig. \ref{fig:ppg}b). 

With the chosen configuration the PPG can generate time bin symbols as short as 1.5 clock cycles: a couple of $\delta t=625$~ps width optical pulses delayed $\Delta T= 1.25$~ns, at a maximum symbol rate of $F_{out}/8$=200~MHz. The specific pattern that defines whether the output symbol is either $\left|\psi_0 \right \rangle$, $\left|\psi_1 \right \rangle$ or $\left|\psi_+ \right \rangle$ is applied in the parallel bus by the FPGA at the symbol rate. It is worth to note that the 8 bit serial pattern allows to code two symbols per serialized word, hence duplicating the symbol rate up to 400~MHz. The output of this pattern generator is later amplified with a MAX3941 (Maxim Integrated) MDA and transferred to the optical domain with the EAM acting on the laser output.

Two amplitude modulators controlled by the FPGA insert a 50\% loss on the $\left|\psi_+ \right \rangle$ states to obtain a uniform photon rate per symbol, and set the relative intensity $\mu_2/\mu_1$ for signal and decoy states. Finally, an additional phase modulator driven by a 50~MHz bandwidth noise generator ensures that the phase is randomized for each symbol. 

The temporal structure of the output signal is set by the FPGA acting on the PPG via the Sync input. Symbols can be clustered in bursts of selectable length and delay, to allow for additional time multiplexing of phase stabilization routines on the detector side, and also to adapt the data stream to different detector dead times. The performance of the device together with a demonstration of quantum state transmission through a simulated optical channel are presented in the following section.

\section{Experimental results}
\label{sec:perform}
\begin{figure}[h!]
\centering\includegraphics[width=10cm]{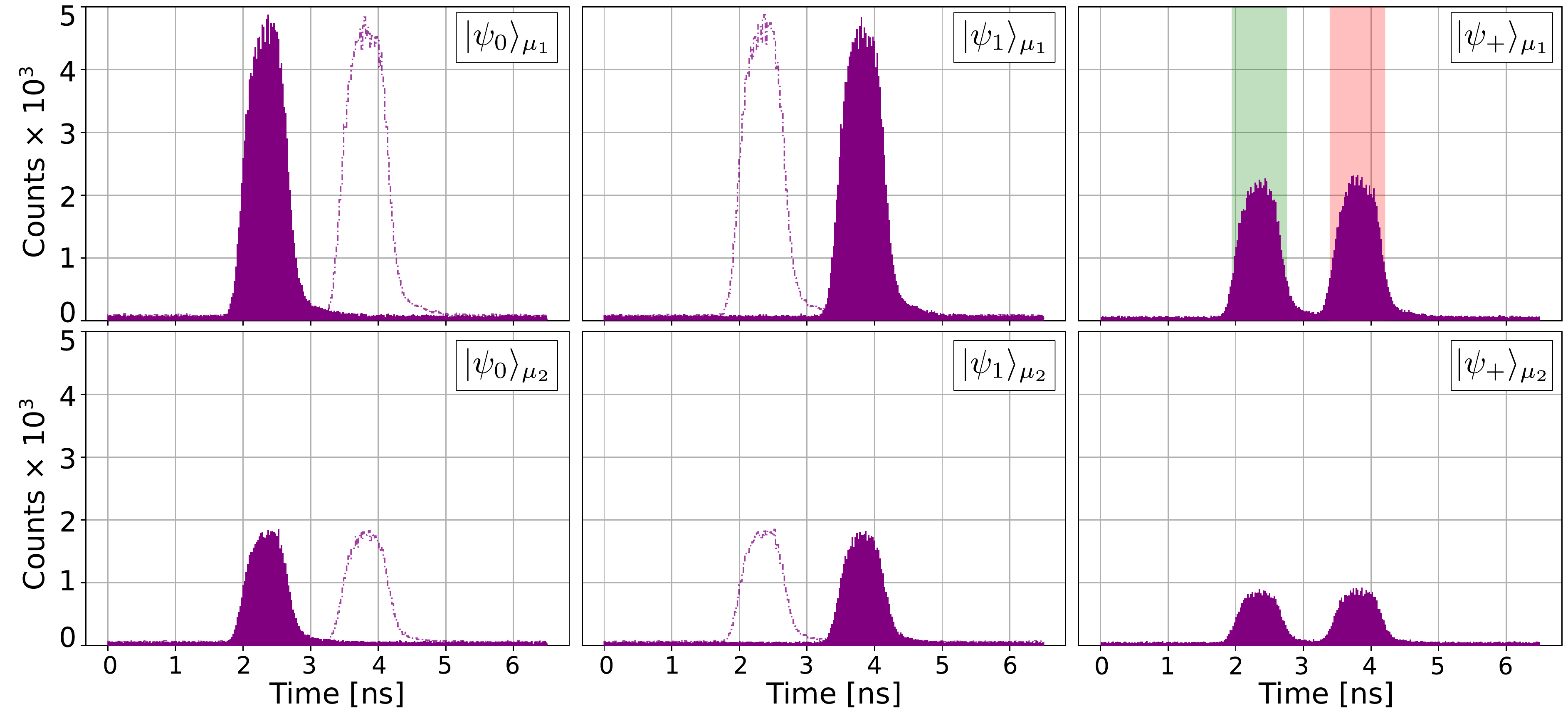}
\caption{Histograms of the measured states proposed in eq.(\ref{eq:estados}). Each state was measured by looking at a fixed symbol within the repeating burst. Valid detections were defined using 0.8~ns windows for each time-bin, shown on the top right panel with green (\emph{early}) and red (\emph{late}) bars. Phantom lines on states of the $Z$ basis show the temporal bin of complementary states.}
\label{fig:estados}
\end{figure}

Histograms of the output optical pulses generated with the transmitter set at a serial clock frequency of 684 MHz and a symbol rate of 5~MHz within the burst are shown in figure \ref{fig:estados}. The ratio $\mu_2/\mu_1\approx0.4$ is close to the optimum for almost any channel length \cite{rusca2018finite}. For the detection of single photons we used an IDQ Id201 InGaAs/InP SPCM with an efficiency of 10\% and a time-to-digital converter (Time Tagger Ultra, Swabian Instruments) with a temporal resolution of 42~ps. The data output was structured in bursts of 20 symbols to account for the 20~$\mu$s detector dead time. A detection window of 20~ns was set on each measurement, for both the $X$ and $Z$ bases.

The transmitter was put under test with two simulated channels of 7~dB and 14~dB total loss, corresponding to fiber optical link lengths of 35~km and 70~km, respectively. For the experimental results presented here, a 300 seconds measurement was performed for each condition, in which an average of 273 million symbols were sent. The mean photon numbers per symbol at the output of the source are $\mu_1=0.50$ and $\mu_2=0.19$ on both cases, and the optimized probabilities of emitting these states are $p_{\mu_1}=0.63$ and $p_{\mu_2}=0.37$. The probability of emitting states of the $Z$ basis was set at $p_Z=0.9$.

\begin{figure}[h!]
\centering\includegraphics[width=8cm]{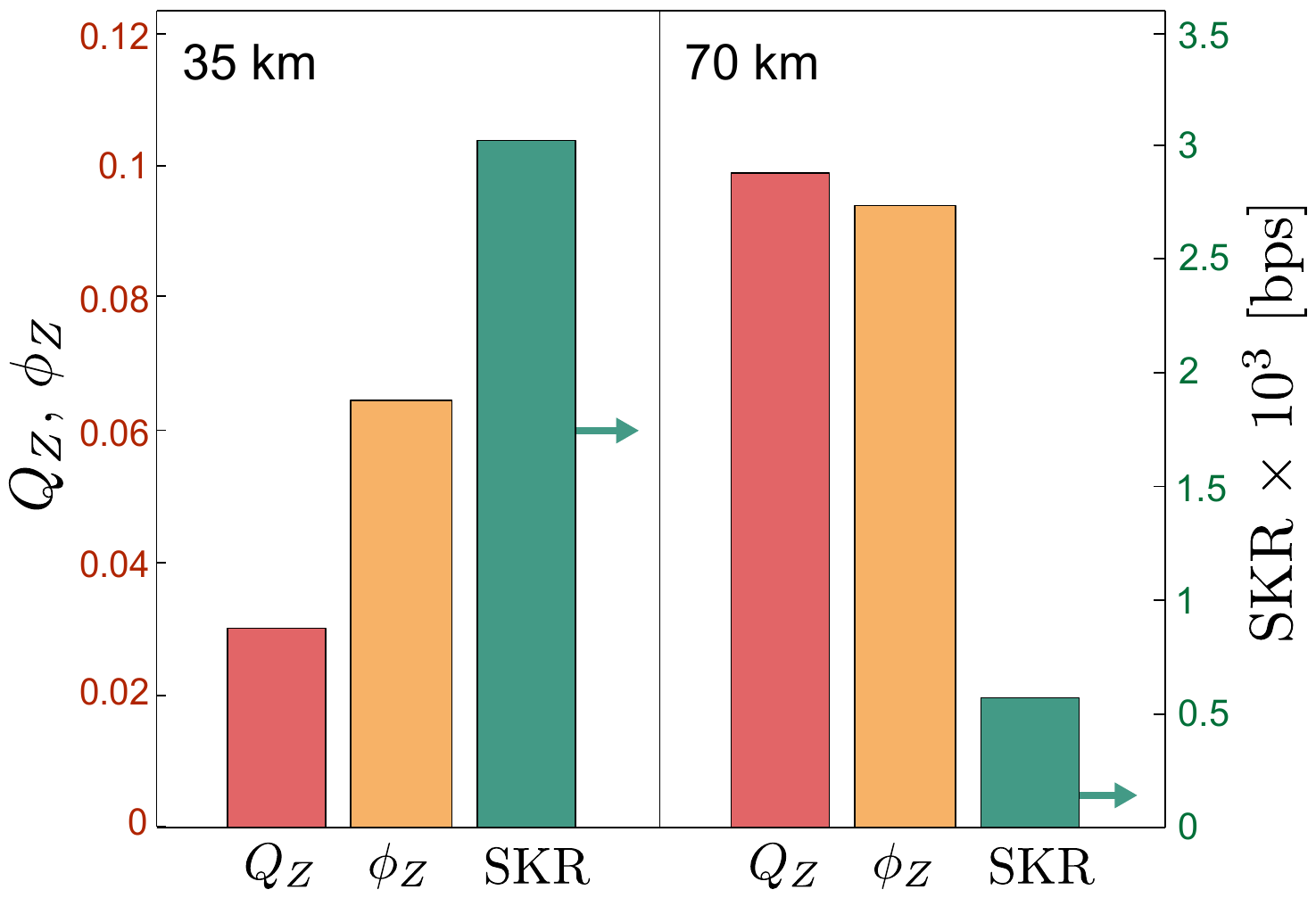}
\caption{Bit and phase error rates obtained for two different simulated lengths of standard single mode fiber, together with the extractable SKR.}
\label{fig:QBIT}
\end{figure}

The bit error in the $Z$ basis $Q_Z$ is obtained in a straightforward way by calculating the ratio between the wrong detections and the total number of detections.  The $\left|\psi_+ \right \rangle$ state is detected and analyzed using an actively stabilized, unbalanced Faraday-Michelson interferometer, which gives a three-pulse output where the interference effect is present on the central pulse. We follow the procedure described in \cite{grande2021adaptable} to estimate the error rate in the $X$ basis, $Q_X$. An upper bound on the phase error rate for states of the $Z$ basis $\phi_Z$ can be obtained using the calculations described in \cite{rusca2018finite} and the QBER estimations $Q_Z$ and $Q_X$. The interferometer was actively stabilized using a modified gradient descent algorithm that maximizes the count rate on the central pulse by acting on a piezoelectric fiber stretcher. Such optimization procedure must be performed every 100 seconds and it is temporally multiplexed between data streams. The bit error rate $Q_Z$ and the phase error rate $\phi_Z$, together with the estimated secret key rates obtained for the two attenuation conditions are summarized in figure \ref{fig:QBIT}. The intrinsic visibility of the interferometer was estimated in 0.98, implying that most of the error in the $X$ basis measurements is due to the detector noise and jitter. The secure key rate can be obtained from the repetition rate of the source and the lower bound of the secret key length in a finite-key scenario, as described in Ref \cite{rusca2018finite}.

For an attenuation equivalent to a fiber length of 35~km we obtain bit and phase error of the order of 3\% and 6\% respectively, and a secure key rate of 3.0~kb per second; when the attenuation is doubled however, the detector dark counts combined with residual imperfections of the $X$ basis detection scheme increase the error rates leading to a reduced SKR of 570 bits per second.

\section{Conclusion}
\label{sec:conclu}

The results presented above demonstrate the use of an electro-optical transmitter for time-bin discrete variable QKD. The configuration can produce the six states needed to implement a three-state protocol with the decoy state method. Fast electrical patterns are delivered with a 8-bit serializer fed by a PLL-synthesized clock, while the timing, gating and low bandwidth synchronizing tasks are implemented within the FPGA board. The specific serializer circuit that was used in this transmitter can deliver output data rates of up to 3.2 GHz. This pattern enables the use of an interferometer with a short delay unbalance for detection in the $X$ basis, by imposing a temporal delay equal to the pulse separation of 625 ps. The arm unbalance can be reduced to an optical fiber segment of only 6.25~cm provided a Michelson-Faraday interferometer is used. The 8-bit variable rate also allows for a flexible output that can accommodate to different technical requirements imposed by the overall channel loss and by the speed and dead time of the detection technology. Furthermore, the serial output can be adapted in a straightforward way to implement high dimensional quantum communication protocols with $d=4$ time bin \emph{qudits} \cite{islam2017provably,vagniluca2020efficient}. The device was tested on a three-state decoy method quantum communication scheme over simulated channels of 35~km and 70~km single mode fiber obtaining secure key yields of $1.1\times 10^{-5}$ and $2.1\times 10^{-6}$ bits per emitted symbol respectively. 
The presented transmitter can be easily combined with a low bandwidth pulse pattern generator or FPGA board to generate time bin quantum states at a high rate for standardized QKD applications.

\begin{backmatter}
\bmsection{Funding}
Content in the funding section will be generated entirely from details submitted to Prism. 

\bmsection{Acknowledgments}
%The section title should not follow the numbering scheme of the body of the paper. Additional information crediting individuals who contributed to the work being reported, clarifying who received funding from a particular source, or other information that does not fit the criteria for the funding block may also be included; for example, ``K. Flockhart thanks the National Science Foundation for help identifying collaborators for this work.'' 

\bmsection{Disclosures}
The authors declare no conflicts of interest.

\bmsection{Data availability} Data underlying the results presented in this paper are not publicly available at this time but may be obtained from the authors upon reasonable request.

\end{backmatter}

%%%%%%%%%%%%%%%%%%%%%%% References %%%%%%%%%%%%%%%%%%%%%%%%%

%%%%%%%%%% If using BibTeX:
\bibliography{opt_trans}

%%%%%%%%%% If preparing manually:
% \begin{thebibliography}{1}
% \newcommand{\enquote}[1]{``#1''}

% \bibitem{Zhang:14}
% Y.~Zhang, S.~Qiao, L.~Sun, Q.~W. Shi, W.~Huang, L.~Li, and Z.~Yang,
%   \enquote{Photoinduced active terahertz metamaterials with nanostructured
%   vanadium dioxide film deposited by sol-gel method,}
%   {\protect\JournalTitle{Optics Express}} \textbf{22}, 11070--11078 (2014).

% \bibitem{Optica}
% {Optica}, \enquote{{Optica Publishing Group},}
%   \url{http://www.opg.optica.org}.

% \bibitem{FORSTER2007}
% P.~Forster, V.~Ramaswamy, P.~Artaxo, T.~Bernsten, R.~Betts, D.~Fahey,
%   J.~Haywood, J.~Lean, D.~Lowe, G.~Myhre, J.~Nganga, R.~Prinn, G.~Raga,
%   M.~Schulz, and R.~V. Dorland, \enquote{Changes in atmospheric consituents and
%   in radiative forcing,} in \enquote{Climate Change 2007: The Physical Science
%   Basis. Contribution of Working Group 1 to the Fourth assesment report of
%   Intergovernmental Panel on Climate Change,}  S.~Solomon, D.~Qin, M.~Manning,
%   Z.~Chen, M.~Marquis, K.~B. Averyt, M.~Tignor, and H.~L. Miler, eds.
%   (Cambridge University Press, 2007).

% \end{thebibliography}

\end{document}